\documentclass[mnsc,nonblindrev]{informs3} % current default for manuscript submission

\OneAndAHalfSpacedXI % current default line spacing
\usepackage{hyperref}
\usepackage{tikz}
\usepackage{xcolor}
\DeclareMathOperator{\E}{\mathbb{E}}

\def\BFA{\boldsymbol{A}}
\def\BFa{\boldsymbol{a}}
\def\BFB{\boldsymbol{B}}
\def\BFb{\boldsymbol{b}}
\def\BFC{\boldsymbol{C}}
\def\BFc{\boldsymbol{c}}
\def\BFD{\boldsymbol{D}}
\def\BFd{\boldsymbol{d}}
\def\BFE{\boldsymbol{E}}
\def\BFe{\boldsymbol{e}}
\def\BFF{\boldsymbol{F}}
\def\BFf{\boldsymbol{f}}
\def\BFG{\boldsymbol{G}}
\def\BFg{\boldsymbol{g}}
\def\BFH{\boldsymbol{H}}
\def\BFh{\boldsymbol{h}}
\def\BFi{\boldsymbol{i}}
\def\BFI{\boldsymbol{I}}
\def\BFJ{\boldsymbol{J}}
\def\BFj{\boldsymbol{j}}
\def\BFK{\boldsymbol{K}}
\def\BFk{\boldsymbol{k}}
\def\BFL{\boldsymbol{L}}
\def\BFl{\boldsymbol{l}}
\def\BFM{\boldsymbol{M}}
\def\BFm{\boldsymbol{m}}
\def\BFN{\boldsymbol{N}}
\def\BFn{\boldsymbol{n}}
\def\BFO{\boldsymbol{O}}
\def\BFo{\boldsymbol{o}}
\def\BFP{\boldsymbol{P}}
\def\BFp{\boldsymbol{p}}
\def\BFQ{\boldsymbol{Q}}
\def\BFq{\boldsymbol{q}}
\def\BFR{\boldsymbol{R}}
\def\BFr{\boldsymbol{r}}
\def\BFS{\boldsymbol{S}}
\def\BFs{\boldsymbol{s}}
\def\BFT{\boldsymbol{T}}
\def\BFt{\boldsymbol{t}}
\def\BFU{\boldsymbol{U}}
\def\BFu{\boldsymbol{u}}
\def\BFV{\boldsymbol{V}}
\def\BFv{\boldsymbol{v}}
\def\BFW{\boldsymbol{W}}
\def\BFw{\boldsymbol{w}}
\def\BFX{\boldsymbol{X}}
\def\BFx{\boldsymbol{x}}
\def\BFY{\boldsymbol{Y}}
\def\BFy{\boldsymbol{y}}
\def\BFZ{\boldsymbol{Z}}
\def\BFz{\boldsymbol{z}}
\def\BFzero{\boldsymbol{0}}
\def\BFone{\boldsymbol{1}}
\def\BFtwo{\boldsymbol{2}}
\def\BFthree{\boldsymbol{3}}
\def\BFfour{\boldsymbol{4}}
\def\BFfive{\boldsymbol{5}}
\def\BFsix{\boldsymbol{6}}
\def\BFseven{\boldsymbol{7}}
\def\BFeight{\boldsymbol{8}}
\def\BFnine{\boldsymbol{9}}
\def\BFalpha{\boldsymbol{\alpha}}
\def\BFbeta{\boldsymbol{\beta}}
\def\BFgamma{\boldsymbol\gamma}
\def\BFGamma{\boldsymbol{\Gamma}}
\def\BFdelta{\boldsymbol{\delta}}
\def\BFDelta{\boldsymbol{\Delta}}
\def\BFepsilon{\boldsymbol{\epsilon}}
\def\BFvarepsilon{\boldsymbol{\varepsilon}}
\def\BFzeta{\boldsymbol{\zeta}}
\def\BFeta{\boldsymbol\eta}
\def\BFtheta{\boldsymbol{\theta}}
\def\BFTheta{\boldsymbol{\Theta}}
\def\BFvartheta{\boldsymbol{\vartheta}}
\def\BFiota{\boldsymbol{\iota}}
\def\BFkappa{\boldsymbol{\kappa}}
\def\BFlambda{\boldsymbol\lambda}
\def\BFLambda{\boldsymbol{\Lambda}}
\def\BFmu{\boldsymbol{\mu}}
\def\BFnu{\boldsymbol{\nu}}
\def\BFxi{\boldsymbol{\xi}}
\def\BFXi{\boldsymbol{\Xi}}
\def\BFpi{\boldsymbol{\pi}}
\def\BFPi{\boldsymbol{\Pi}}
\def\BFrho{\boldsymbol{\rho}}
\def\BFvarrho{\boldsymbol{\varrho}}
\def\BFsigma{\boldsymbol{\sigma}}
\def\BFSigma{\boldsymbol{\Sigma}}
\def\BFtau{\boldsymbol{\tau}}
\def\BFupsilon{\boldsymbol\upsilon}
\def\BFUpsilon{\boldsymbol\Upsilon}
\def\BFphi{\boldsymbol{\phi}}
\def\BFvarphi{\boldsymbol{\varphi}}
\def\BFPhi{\boldsymbol{\Phi}}
\def\BFchi{\boldsymbol{\chi}}
\def\BFpsi{\boldsymbol{\psi}}
\def\BFPsi{\boldsymbol{\Psi}}
\def\BFomega{\boldsymbol{\omega}}
\def\BFOmega{\boldsymbol{\Omega}}
\definecolor{bluea}{RGB}{192,188,248}

\usepackage{pgfplots}
\pgfplotsset{compat=newest}
\usepgfplotslibrary{polar}

\usepackage{endnotes}

\let\footnote=\endnote
\let\enotesize=\normalsize
\def\notesname{Endnotes}%
\def\makeenmark{$^{\theenmark}$}
\def\enoteformat{\rightskip0pt\leftskip0pt\parindent=1.75em
  \leavevmode\llap{\theenmark.\enskip}}

\usepackage{natbib}
 \bibpunct[, ]{(}{)}{,}{a}{}{,}%
 \def\bibfont{\small}%
 \usepackage{makecell}

\usepackage{algorithm,bm}
\usepackage{algorithmicx}
\usepackage{algpseudocode}
\algnewcommand{\algorithmicand}{\textbf{and }}
\algnewcommand{\algorithmicor}{\textbf{or }}
\algnewcommand{\OR}{\algorithmicor}
\algnewcommand{\AND}{\algorithmicand}
\makeatletter
\newenvironment{breakablealgorithm}
  {% \begin{breakablealgorithm}
  \begin{center}
     \refstepcounter{algorithm}% New algorithm
     \hrule height.8pt depth0pt \kern2pt% \@fs@pre for \@fs@ruled
     \renewcommand{\caption}[2][\relax]{% Make a new \caption
      {\raggedright\textbf{\ALG@name~\thealgorithm} ##2\par}%
      \ifx\relax##1\relax % #1 is \relax
         \addcontentsline{loa}{algorithm}{\protect\numberline{\thealgorithm}##2}%
      \else % #1 is not \relax
         \addcontentsline{loa}{algorithm}{\protect\numberline{\thealgorithm}##1}%
      \fi
      \kern2pt\hrule\kern2pt
     }
  }{% \end{breakablealgorithm}
     \kern2pt\hrule\relax% \@fs@post for \@fs@ruled
  \end{center}
  }

\usepackage{makecell}

\usepackage{ulem}
\usepackage{booktabs}
\usepackage{multirow}
\usepackage{comment}

\usepackage{hyperref}

\usepackage{xcolor}
\hypersetup{
    colorlinks,
    linkcolor={red!50!black},
    citecolor={blue!50!black},
    urlcolor={blue!80!black}
}
\TheoremsNumberedThrough     % Preferred (Theorem 1, Lemma 1, Theorem 2)
\ECRepeatTheorems

\EquationsNumberedThrough    % Default: (1), (2), ...
\newenvironment{assumption'}[1]
  {\renewcommand{\theassumption}{\arabic{assumption}$'$}%
   \addtocounter{assumption}{-1}%
   \begin{assumption}}
  {\end{assumption}}

\usepackage{amsmath}
\usepackage{amsfonts}
\usepackage{amssymb}
\usepackage{dsfont}
\newenvironment{alg'}[1]
  {\renewcommand{\thealgorithm}{\ref{#1}$'$}%
   \addtocounter{algorithm}{-1}%
   \begin{breakablealgorithm}}
  {\end{breakablealgorithm}}

\newenvironment{lemma'}[1]
  {\renewcommand{\thelemma}{\ref{#1}$'$}%
   \addtocounter{lemma}{-1}%
   \begin{lemma}}
  {\end{lemma}}

\begin{document}
%%%%%%%%%%%%%%%%

%\setcounter{page}{1}
%\DOI{10.1287/xxxx.0000.0000}%

% \RUNAUTHOR{A,B and C}

\RUNTITLE{RT}

% \TITLE{Bundling Information Goods of Decreasing Value}
\TITLE{A Preliminary Study on Accelerating Simulation Optimization with GPU Implementation}


\ARTICLEAUTHORS{%
\AUTHOR{Jinghai He, Haoyu Liu, Yuhang Wu, Zeyu Zheng, Tingyu Zhu}
\AFF{Department of Industrial Engineering and Operations Research\\
University of California, Berkeley}
%\AUTHOR{CD}
%\AFF{Institute2}
} % end of the block

\ABSTRACT{%
We provide a preliminary study on utilizing GPU (Graphics Processing Unit) to accelerate computation for three simulation optimization tasks with either first-order or second-order algorithms. Compared to the implementation using only CPU (Central Processing Unit), the GPU implementation benefits from computational advantages of parallel processing for large-scale matrices and vectors operations. Numerical experiments demonstrate  computational advantages of utilizing GPU implementation in simulation optimization problems, and show that such advantage comparatively further increase as the problem scale increases. 
}%

%\KEYWORDS{deterministic inventory theory; infinite linear programming duality;
%  existence of optimal policies; semi-Markov decision process; cyclic schedule}

% \KEYWORDS{Graphics Processing Unit, simulation optimization, parallel computing} 
%
% \HISTORY{This paper was first submitted on }

\maketitle
%%%%%%%%%%%%%%%%%%%%%%%%%%%%%%%%%%%%%%%%%%%%%%%%%%%%%%%%%%%%%%%%%%%%%%

% Samples of sectioning (and labeling) in OPRE
% NOTE: (1) \section and \subsection do NOT end with a period
%       (2) \subsubsection and lower need end punctuation
%       (3) capitalization is as shown (title style).
%
%\section{Introduction.}\label{intro} %%1.
%\subsection{Duality and the Classical EOQ Problem.}\label{class-EOQ} %% 1.1.
%\subsection{Outline.}\label{outline1} %% 1.2.
%\subsubsection{Cyclic Schedules for the General Deterministic SMDP.}
%  \label{cyclic-schedules} %% 1.2.1
%\section{Problem Description.}\label{problemdescription} %% 2.

% Text of your paper here

\label{sec:intro}

\section{Introduction}
Simulation optimization (SO) generally refers to optimization in the setting where the objective function $f(x)$ and/or the constraints $\Theta$ involves uncertainty and cannot be directly analytically evaluated and can only be evaluated through simulation experiments; see \cite{fan2024review} for a recent review on simulation optimization. A general simulation optimization problem can be represented by the follows \citep{jian2015introduction},
\begin{align}\label{Eq: Simopt}
\min_{x \in \Theta}\, &\mathbb{E}[f(x, \xi)], \\
\text{where } \Theta = &\{ x : \mathbb{E}[g(x, \xi)] \geq 0\} \nonumber
\end{align}
where \( x \) are the decision variables, $\xi$ are the random variables representing the randomness in the system, where \( f(x,\xi) \) denotes one stochastic realization of objective via simulation and $g(x,\xi)$ denotes one stochastic realization of the constraint via simulation. We refer to  \cite{hong2009brief}, \cite{fu2015handbook}, \cite{peng2023simulation} and \cite{fan2024review} for more detailed review.

% Based on the structure of feasible region $\Theta$ the simulation optimization problem can be categorized into three categories: 1) ranking and selection where $\Theta$ is finite set; 2) continuous simulation optimization where $\Theta$ is a subset of $\mathbb{R}^d$; and 3) Discrete simulation optimization where $\Theta$ is integer lattice. 

Classical implementation of simulation optimization algorithms on computers has mainly been using CPU (Central Processing Unit) by default, without a specialized use of GPU (Graphics Processing Unit). Research on parallelization and synchronization of simulation optimization algorithms has also largely been designed and implemented for CPU-based computation, or at least not specializing the use of GPU. Recent developments in the computational tools (for broad purposes) have indicated that the use of GPUs may provide specialized advantages in acceleration, if used appropriately.  

In this work, we consider three sub-classes of simulation optimization problems and investigate the use of GPUs (compared to not using GPUs) to accelerate the algorithms' computational speed while maintaining a similar level of solution accuracy. Specifically, we focus on leveraging the superior capabilities of GPUs for conducting large-scale matrices and vectors operations and parallel processing to  enhance the efficiency and performance of simulation optimization algorithms.

\subsection{Background}

The Graphics Processing Unit (GPU), originally designed for accelerating graphics rendering, has evolved into a cornerstone for parallel processing tasks. Unlike Central Processing Units (CPUs) that excel in sequential task execution with few cores, GPUs feature thousands of smaller cores optimized for handling multiple operations in parallel, making them highly efficient for parallelizable tasks \citep{kirk2007nvidia,owens2008gpu}.

In recent-year development of machine learning and deep learning, GPU with their parallel processing prowess, have significantly accelerated the training and inference processes of complex neural network architectures. This acceleration is particularly crucial for handling the vast amounts of data and the computationally intensive tasks inherent in computer vision \citep{he2017mask,gu2018recent}. The development of transformer-based deep learning models further underscores the power of GPUs in  facilitating the exploration of more advanced models. The inherent parallelism of GPUs makes them ideal for this task, enabling the rapid processing of the large-scale matrix operations that are central to transformers \citep{fei2017high}.

% The advantage of GPUs in parallel computing is not confined to machine learning (ML) but also extends to other domains requiring extensive computation, such as biology \citep{phillips2020scalable}, physics \citep{makoviychuk2021isaac}, chemistry \citep{desai2021direct} and optimization \citep{lu2023cupdlp}. The utility of GPUs in these fields leverages the capability for parallel computation to manage high-dimensional calculations and to simulate experimental scenarios efficiently.

In our work, we attempt to study the computational prowess of GPUs within the domain of simulation optimization, which often requires continuous simulations and sampling as well as intensive matrix computations. Our work is connected to the large-scale simulation optimization literature. When the size of the SO problem gets larger, the feasible region may grow exponentially with the dimension of the decision variable. This curse of dimensionality leads to computational challenges, such as simulating exponentially more observations to estimate the objective function, low rate of convergence \citep{gao2020towards,wang2023gaussian} and smoothness problem of the objective function \citep{ding2021high,erdogdu2021convergence}. To tackle these challenges, various approaches \citep{kandasamy2015high,zhang2021efficient,rolland2018high,xu2013empirical,gao2015note,pearce2022bayesian,hong2022solving} are designed for more efficient computation and estimation. We refer to \cite{fan2024review} for more thorough review on these methods. We also refer to \cite{eckman2023simopt} for a broad testbed of simulation optimization problems. \cite{l2017random} comprehensively studied the use of GPU for random number generation. 
\cite{peng2024} considered the use of GPU to accelerate policy evaluation in a general reinforcement learning problem setting.  Another branch of methods in solving high-dimensional and large-scale simulation optimization problem is using parallelization \citep{zhang2024sample,ni2017efficient,luo2015fully}. Different from the mentioned works, our work specifically focuses on the simulation optimization tasks of which the main computation can be completed through matrix operations and vectorization. 

% his property make the tasks suitable for acceleration via GPU computation.

\subsection{Organization and summary}
The rest of our work is organized as follows. In Section 2, we briefly overview the architecture of Graphics Processing Units (GPUs), and the mechanisms behind computation acceleration for simulation optimization with GPU implementation.
In Section 3, we present the formulations of three sub-classes of optimization tasks: portfolio optimization, the multi-product Newsvendor problem, and a binary classification problem. The simulation optimization algorithms involved to address these problems include the Frank-Wolfe algorithm and the stochastic quasi-Newton algorithm.

In Section 4, we design and implement the numerical experiments via the use of GPU for the tasks and algorithms introduced in Section 3.  We implement the algorithms on GPUs with JAX library and conduct a comparative analysis against their performance on CPUs across various task sizes (ranging from 100 to $1\times 10^6$ decision variables). Our findings show that, when executed on GPUs, the algorithms operate between three to five times faster than their CPU counterparts, while maintaining similar solution accuracy and convergence. From the experiments, this running time difference between GPU version and CPU version becomes increasingly pronounced in larger-scale problems. Section 5 draws conclusions on our preliminary study of GPU implementation for some simulation optimization problems and the limitations of our work.

\section{GPU for Computation Acceleration}
\subsection{GPU Architecture}
\begin{figure}[htbp]
    \centering
    \includegraphics[scale = 0.75]{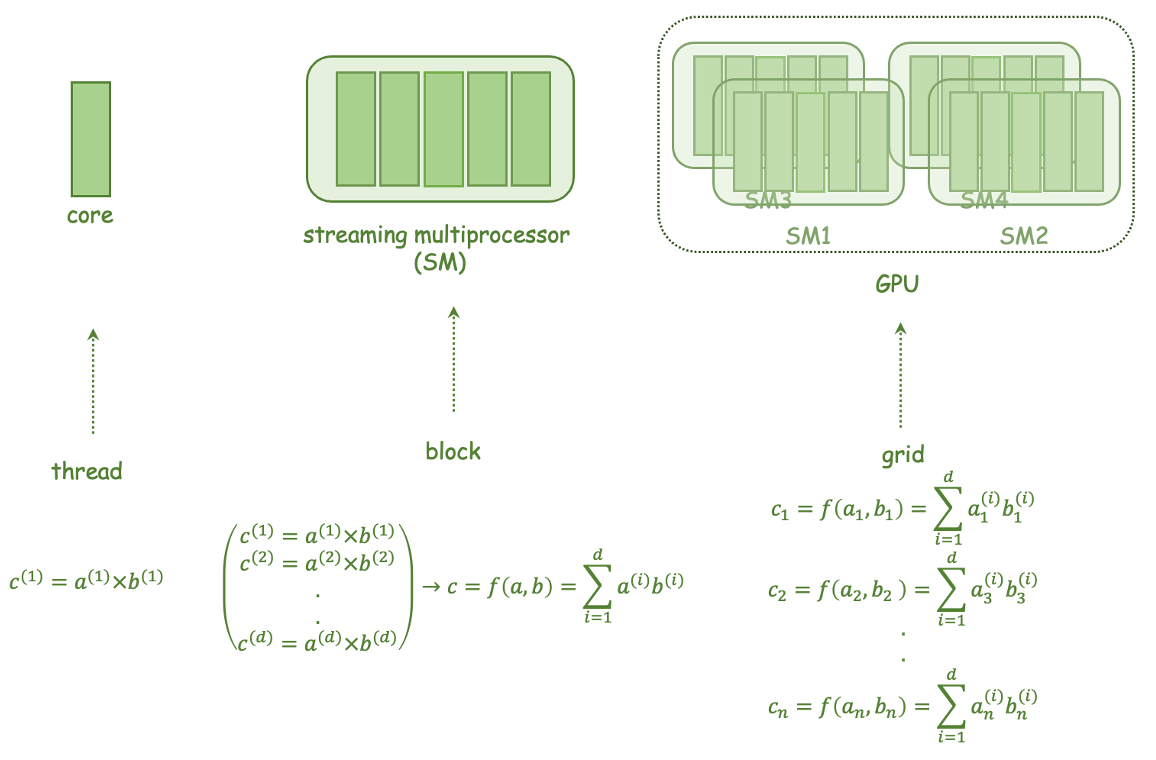}
    \caption{GPU Architecture}
    \label{fig:cuda}
\begin{flushleft}
\small
\textbf{Notes:}
This figure presents the architecture of a GPU, and how each parts work parallel in a naive example for calculating inner product of two vectors. A thread refers to the calculation of $c^{(1)}=a^{(1)}\times b^{(1)}$, and is processed by a core. A block refers to the parallel operation of $d$ threads, and is processed by a SM. A grid refers to the parallel operation of $n$ blocks, and is processed by a GPU. 
\end{flushleft}
\end{figure}

We use the most widely used GPU Architecture: Compute Unified Device Architecture (cuda) shown in Figure \ref{fig:cuda} as example to introduce the structure of GPU and how it works for vectorization and parallel computing.

A modern GPU typically has more than thousands of cores, with each core being the most basic unit designed to carry out arithmetic operations. The arithmetic operations are referred to as `thread' from the task perspective. The cores of a GPU are first grouped into larger units known as streaming multiprocessors (SMs), and then further organized on a grand scale to form a GPU. From the task perspective, a task processed by a SM is referred to as a `block'; examples include vector operations. Tasks supported by the GPU, such as parallel computing of multiple functions, are referred to as `grid'. The computational framework of GPUs follows the `Single Instruction, Multiple Data' (SIMD) model, where multiple threads carry out the same operation but with different data sets in a block. This SIMD model naturally benefit the operations in vectorization (vector/matrix) forms.

\subsection{Acceleration of Simulation Optimization with GPU} \label{sec:acceleration}
For simulation optimization tasks, especially those on a large scale, two main computational bottlenecks emerge: the first involves operations such as vector-vector, matrix-vector, and matrix-matrix multiplications, while the second pertains to the sampling process required for estimating objective values or gradients, which demands numerous iterations, often ranging from tens to hundreds, for each computational step. The architecture of GPU is fundamentally aligned with the principles of parallel computing, making GPUs an effective platform for  handling both types of operations. This parallel computing capability not only facilitates rapid execution of complex matrix and vector operations but also potentially accelerates the sampling process by enabling simultaneous execution of multiple iterations. 

% For vector/matrix-based operations, consider the computation of the inner product between two vectors, \(\mathbf{a}\) and \(\mathbf{b}\), defined as \(\mathbf{a} \cdot \mathbf{b} = \sum_{i=1}^d a_ib_i\), where \(\mathbf{a}, \mathbf{b} \in \mathbb{R}^d\), and \(a_i, b_i\) denote the elements at the \(i^{th}\) index of vectors \(\mathbf{a}\) and \(\mathbf{b}\), respectively. In a CPU-based computing environment, this operation is typically executed sequentially, processing each element pair \(a_ib_i\) in succession. Conversely, the parallel computation framework of GPUs facilitates the simultaneous computation of each product \(a_ib_i\) across multiple processing units, thereby significantly reducing the total computation time. This parallel processing capability is particularly advantageous in scenarios where the dimensionality, \(d\), is substantial, highlighting the superiority of GPUs in managing high-dimensional data and complex operations with remarkable efficiency.

In the lower section of Figure~\ref{fig:cuda}, we briefly illustrate the operational complexities of the GPU architecture that facilitate parallel computation, exemplified by the calculation of vector inner products. Each core within the Streaming Multiprocessor (SM) executes a thread responsible for multiplying individual vector elements. As illustrated, a thread computes the product $(a_i \cdot b_i)$, with these operations occurring concurrently across all threads within a block. Within each block, inner products are computed as multiple cores work in parallel to derive the final results. Concurrently, multiple inner product functions can be executed within each grid.

Concerning the sampling process used for estimating objective values or gradients, traditional CPU-based systems usually conduct these operations sequentially, processing each sample individually. In contrast, GPUs leverage their parallel processing capabilities to handle multiple sampling operations simultaneously. Each SM in Figure~\ref{fig:cuda} executes a single sampling operation for an objective function, and multiple SMs can operate in parallel to sample different pathways concurrently. 

% The dual benefits of GPUs—accelerating both algebraic and stochastic computations—substantially aid in managing large-scale simulation optimization challenges, enabling researchers and practitioners to achieve faster convergence to optimal solutions with enhanced accuracy. 

\section{Three Simulation Optimization Tasks} In this section, we consider three simulation optimization tasks that all fall into the following formulation subject to deterministic linear constraints, formulated as follows:

\begin{equation}
\begin{aligned}
    \min_{\theta\in\mathbb{R}^d} f(\theta) &= \E F(X;\theta)\\
    \text{s.t. } A\theta&=b.
\end{aligned}    
\end{equation}

Many algorithms have been developed for solving this class of optimization problem. In this work, we employ the Frank-Wolfe algorithm on two specific tasks: the mean-variance portfolio optimization and the multi-product Newsvendor problem. Subsequently, we shift our focus to second-order optimization, where we implement a stochastic quasi-Newton algorithm for tackling a binary classification problem. Each of these applications is discussed in further detail below. The three algorithms under consideration predominantly depend on vectorization computations and necessitate the estimation of sample gradients. 

% These processes are inherently suited for acceleration via GPU-based parallel computing, as discussed in Section \ref{sec:acceleration}.

\subsection{Task1: Mean-variance Optimization}
We first consider a general mean-variance optimization problem given by
\begin{equation}
\label{task 1}
    \begin{aligned}
    \min_{\bm{w}\in\mathbb{R}^d} f(\bm{w}) &=  \frac{1}{2} \mathrm{Var}[\bm{w}^\top \bm{R}] - \E[\bm{w}^\top \bm{R}]\\
    &= \frac{1}{2} \bm{w}^\top \mathrm{Cov}[\bm{R}]\bm{w} - \bm{w}^\top\E[\bm{R}]\\
    \text{s.t. } \bm{w}^\top\bm{1}&\leq 1,\\
    \bm{w}&\geq 0,
\end{aligned}
\end{equation}
where $\bm{R}$ follows some known distribution, say $\bm{R}\sim \mathcal{N}(\bm{\mu},\bm{\Sigma})$. Besides, we assume that $\E[\bm{R}]$ and $\mathrm{Cov}[\bm{R}]$ are not explicitly given, but we are allowed to draw samples from the distribution and approximate the mean and covariance matrix accordingly. Specifically, the approximated objective function is
\begin{equation}
    \hat{f}(\bm{w}) = \frac{1}{N-1}\sum_{i=1}^N \bm{w}^\top (\bm{R}_i-\bar{\bm{R}})(\bm{R}_i-\bar{\bm{R}})^\top \bm{w} - \bm{w}^\top\bar{\bm{R}},
\end{equation}
where
\begin{equation}
    \bar{\bm{R}} = \frac{1}{N}\sum_{i=1}^N \bm{R}_i
\end{equation}
and $\bm{R}_i,i=1,2,\cdots,N$ are i.i.d. samples drawn from the target distribution. We apply the Frank-Wolfe algorithm to solve the problem, and we re-sample the $\bm{R}_i$'s after every $M$ iterations. The algorithm is presented in Algorithm~\ref{mean-variance optimization: frank-wolfe algorithm}.

\begin{algorithm}
\caption{Frank-Wolfe Algorithm for Mean-variance Optimization}
\label{mean-variance optimization: frank-wolfe algorithm}
\begin{algorithmic}[1]
\State \textbf{Input:} Distribution of $\mathcal{D}(\bm{R})$, constraint set $W=\{\bm{w}|\bm{w}^\top\bm{1}\leq 1, \bm{w}\geq 0\}$, objective function $f$, starting point $\bm{w}_0 \in W$, number of iterations between resampling $M$, number of epochs $K$.
\State \textbf{Output:} Optimal $\bm{w}$
\State Initialize $\bm{w}_0$
\For{$k = 0$ \textbf{to} $K-1$}
    \State Resample $\mathbf{R}_i,i=1,\cdots,N$ from $\mathcal{D}$ \Comment{Sample size can be $N_k$ which is adapted to $k$}
    \For{$m = 0$ \textbf{to} $M-1$}
        \State Compute gradient $\nabla \hat{f}(\bm{w}_m)$
        \State Solve the linear subproblem $\bm{s}_m = \arg\min_{\bm{s} \in W} \bm{s}^\top\nabla \hat{f}(\bm{w}_m)$
        \State Compute step size $\gamma_m = \frac{2}{kM+m+2}$
        \State Update $\bm{w}_{m+1} = \bm{w}_m + \gamma_m(\bm{s}_m - \bm{w}_m)$
    \EndFor
    \State $\bm{w}_0 \leftarrow \bm{w}_{M}$
\EndFor
\State \textbf{return} $\bm{w}_0$
\end{algorithmic}
\end{algorithm}

\subsection{Task2: Multi-product Newsvendor Problem}
We consider a multi-product constrained Newsvendor problem with independent product demands \citep{niederhoff2007using}. The decision maker is interested in jointly determining the stock level $x_j$ for product $j=1,\cdots,N$ to satisfy overall customer demand. For each product $j$, the customer demand is characterized by a stochastic distribution with cdf $\Phi_j$ and pdf $\phi_j$. The unit cost of product $j$ is $k_j$; the holding cost per unit is $h_j$ ($h_j<0$ means scrap value); and the selling value per unit is $v_j$. Thus, the expected cost objective for product $j$ is given by
\begin{equation}
    f_j(x_j)=k_j x_j + h_j\int_0^{x_j}(x_j-\xi)\phi_j(\xi)\mathrm{d}\xi + v_j \int_{x_j}^\infty(\xi-x_j)\phi_j(\xi)\mathrm{d}\xi.
\end{equation}
The stocking quantities are subject to some ex-ante linear constraints which represents the budget constraint for resources. Suppose that there are $M$ resources to be considered, with the constraint level for resource $i$ being $C_i$. The resource requirement for product $j$ and resource $i$ is $c_{i,j}$. For simplicity, we denote the technology matrix as $A^{M \times N}=(c_{i,j})$ and the vector of constraints as $C^{M\times 1}=(C_1,\cdots,C_M)^\top$. Non-negativity of $x_j$ is also assumed. Therefore, the decision-making problem takes the form
\begin{equation}
    \begin{aligned}
        \min_{x_1,\cdots,x_N} f(x)&=\sum_{j=1}^N f_j(x_j)\\
        \text{s.t. } Ax&\leq C\\
        x_j&\geq 0, j=1,\cdots,N.
    \end{aligned}
\end{equation}
For now, we can assume that the demand for each product $j$ follows a normal distribution $\mathcal{N}(\mu_j,\sigma_j^2)$, and the probability of negative demand is negligible as long as $\sigma_j$ is small (compared with $\mu_j$).

Since the total cost is a sum of $N$ separable convex functions, the gradient $\nabla f(x)$ follows
\begin{equation*}
    \nabla f(x)^\top = \left(f_1^\prime(x_1),\cdots,f_N^\prime(x_N)\right)^\top,
\end{equation*}
with
\begin{equation}
    f_j^\prime(x_j) = k_j-v_j+(h_j+v_j)\Phi(x_j).
\end{equation}

Presume that in a situation we do not have closed-form representation for $\Phi(\cdot)$, and we would approximate that in $f_j^\prime$ by Monte Carlo simulation. The approximated gradient is given by
\begin{equation}
\label{newsvendor: approximated gradient}
    \hat{f_j}^\prime(x_j) = k_j-v_j+(h_j+v_j)\frac{1}{S_j}\sum_{j=1}^{S_j}\mathbb{I}\{d_j^{(s)}\leq x_j\},
\end{equation}
where $d_j^{(s)},s=1,\cdots,S_j$ are $S_j$ i.i.d. samples from the demand distribution. The algorithm is provided in Algorithm~\ref{newsvendor: frank-wolfe algorithm}, which is similar with Algorithm~\ref{mean-variance optimization: frank-wolfe algorithm}. For illustration purposes, we simply use Gaussian distribution in the numerical experiments. 

\begin{algorithm}
\caption{Frank-Wolfe Algorithm for Newsvendor Problem}
\label{newsvendor: frank-wolfe algorithm}
\begin{algorithmic}[1]
\State \textbf{Input:} Demand distribution for each product $j$, constraint set $X=\{x|Ax\leq C, x\geq 0\}$, objective function $f$, starting point $x_0 \in X$, number of iterations between resampling $M$, number of epochs $K$.
\State \textbf{Output:} Optimal $x$
\State Initialize $x^{(0)}$
\For{$k = 0$ \textbf{to} $K-1$}
    \State For each $j$, resample $d_j^{(s)}$,$s=1,\cdots,S_j$ from $\mathcal{N}(\mu_j,\sigma_j^2)$
    \For{$m = 0$ \textbf{to} $M-1$}
        \State Compute gradient $\nabla \hat{f}(x^{(m)})=\left(\hat{f}_1(x_1^{(m)}),\cdots,\hat{f}_N(x_N^{(m)})\right)^\top$ according to (\ref{newsvendor: approximated gradient})
        \State Solve the linear subproblem $s^{(m)} = \arg\min_{s \in X} s^\top\nabla \hat{f}(x^{(m)})$
        \State Compute step size $\gamma_m = \frac{2}{kM+m+2}$
        \State Update $x^{(m+1)} = x^{(m)} + \gamma_m(s^{(m)} - x^{(m)})$
    \EndFor
    \State $x^{(0)} \leftarrow x^{(M)}$
\EndFor
\State \textbf{return} $x^{(0)}$
\end{algorithmic}
\end{algorithm}
\subsection{Task3: Binary Classification Problem}

In this section, we consider second-order quasi-Newton-type algorithms. Specifically, we consider a binary classification problem from \cite{byrd2016stochastic}, given as
\begin{equation}
    \min_\omega F(\omega)=-\frac{1}{N}\sum_{i=1}^N z_i\log(c(\omega;x_i))+(1-z_i)\log(1-c(\omega;x_i)),     
\end{equation}
where 
\begin{equation}
    c(\omega,x_i)=\frac{1}{1+\exp(-x_i^T\omega)},~x_i\in\mathbb{R}^n,~\omega\in\mathbb{R}^n,
\end{equation}
and $z_i\in\{0,1\}$. To explain, $x_i$ ($i=1,2,\ldots,N$) denote the feature values of each data point $i$, and $z_i$ is the corresponding classification label. The $n$ data points may come from sample collection or Monte Carlo simulation, depending on different application cases. The objective function is minimized (w.r.t. $\omega$) to derive a function $c_{\omega^*}(x_i)$ that best predicts the label $z_i$ of the sample $x_i$.

% The stochasticity of the problem lies in sampling: every updating step of $\omega$ is calculated based on $b\ll N$ (or $b_H\ll N$) randomly sampled points from the dataset $\{(x_i,z_i)\}_{i=1}^N$.

To solve the problem, we apply the GPU implementation of the stochastic quasi-Newton method (SQN) provided in \cite{byrd2016stochastic} as given in Algorithm \ref{alg:SQN}. Specifically, we use a mini-batch stochastic gradient based on $b=|\mathcal{S}|$ sampled pairs of $(x_i,z_i)$, yielding the following estimate of gradient
\begin{equation}
  \label{eq:gradient_SQN}  \widehat{\nabla}F(\omega)=\frac{1}{b}\sum_{i\in\mathcal{S}}\nabla f(\omega; x_i,z_i),
\end{equation}
where $f(\omega; x_i,z_i)=z_i\log(c(\omega;x_i))+(1-z_i)\log(1-c(\omega;x_i))$.
Further, let \begin{equation}
\label{eq:Hessian_AQN}
\widehat{\nabla}^2 F(\omega)=\frac{1}{b_H}\sum_{i\in\mathcal{S}_H}\nabla^2f(\omega;x_i,z_i)
\end{equation}
be a sub-sampled Hessian, where $\mathcal{S}_H\in\{1,\ldots,N\}$ is also randomly sampled, and $b_H=|\mathcal{S}_H|$. The rest of the details are given in Algorithm \ref{alg:SQN}.

\begin{algorithm}[htbp]
\caption{SQN Algorithm for the Classification Problem}
\label{alg:SQN}
\begin{algorithmic}[1]
\State \textbf{Input:} Step parameter integer $L>0$, memory integer $M>0$, step length parameter $\beta>0$, sample size parameters $b$, $b_H$; initial point $\omega^1$
\State \textbf{Output:} Optimal $\omega$
\State Set $t=-1$, $\bar{\omega}_{t}=0$.\Comment{$t$ records number of correction pairs currently computed}
\For{$k = 1,2,\ldots$ }
    \State Choose a sample set $\mathcal{S}\in\{1,2,\ldots,N\}$
    \State Calculate stochastic gradient $\hat{\nabla}F(\omega_t)$ by \eqref{eq:gradient_SQN}.
    \State $\bar{\omega}_t=\bar{\omega}_t+\omega^k$, $\alpha_k=\beta/k$
    \If{$k\leq 2L$}
        \State $\omega^{k+1}=\omega^k-\alpha^k\hat{\nabla}F(\omega^k)$\Comment{Stochastic gradient iteration}
    \Else
        \State $\omega^{k+1}=\omega^k-\alpha^k H_t \hat{\nabla}F(\omega^k)$, where $H_t$ is calculated by Algorithm \ref{alg:Hessian}.
    \EndIf
    \If{$\mod(k,L)=0$}\Comment{Compute correction pairs every $L$ iterations}
    \State $t=t+1$
    \State $\bar{\omega}_t=\bar{\omega}_t/L$.
        \If{$t>0$}
        \State Choose a sample $\mathcal{S}_H\in\{1,\cdots ,N\}$ to compute $\hat{\nabla}^2 F(\bar{\omega}_t)$ by \eqref{eq:Hessian_AQN}.
        \State Compute $s_t=(\bar{\omega}_t-\bar{\omega}_{t-1})$, $y_t=\hat{\nabla}^2 F(\bar{\omega}_t)(\bar{\omega}_t-\bar{\omega}_{t-1})$.\Comment{Correction pairs}
        \EndIf
        \State $\bar{\omega}_t=0$
    \EndIf
    
\EndFor
\State \textbf{return} $\omega^k$
\end{algorithmic}
\end{algorithm}
\begin{algorithm}[htbp]
\caption{Hessian Updating}
\label{alg:Hessian}
\begin{algorithmic}
\State \textbf{Input:} Updating counter $t$, memory integer $M>0$, and correction pairs $(s_j,y_j)$ where $j =t-\tilde{m}+1,\ldots,t$ and $\tilde{m}=\min\{t, M\}$ \Comment{All come from Algorithm \ref{alg:SQN}}
\State \textbf{Output:} new matrix $H_t$
\State set $H = (s^T_ty_t)/(y^T_t y_t)I$, where $s_t$ and $y_t$ are computed from Algorithm \ref{alg:SQN}.
\For{$j=t-\tilde{m}+1,\ldots,t$}
\State $\rho_j=1/y_j^Ts_j$
\State Apply BFGS formula:
\begin{equation*}
    H \leftarrow(I-\rho_js_jy^T_j )H(I-\rho_jy_js^T_j) +\rho_js_js^T_j
\end{equation*}
\EndFor
\State \textbf{return} $H_t\leftarrow H$
\end{algorithmic}
\end{algorithm}

\section{Numerical Results}
In this section, we study the numerical performance of the designed algorithm implemented on GPU for three mentioned simulation optimization tasks. In particular, we compare
the implemented simulation optimization algorithms on GPU with their counterpart versions on CPU for the computation time and accuracy.

\subsection{Experimental setup}

We use JAX \citep{jax2018github} library for implementation. We executed the same algorithm for each task using identical parameters on both CPU and GPU (with JAX) across problems of varying sizes to illustrate the performance discrepancies between CPU and GPU at different scales. Our primary focus was on the computational time metric, which we estimated based on the computation time required for comparable iterations of the algorithms on both platforms. Additionally, we evaluated the accuracy and convergence metrics, defined by the relative squared error (RSE) in each iteration relative to the final objective values. The parameters of each task are listed below.

For Task 1, we consider the mean-variance optimization problem of asset sizes of $5\times 10^2,5\times 10^3,1\times 10^4,5\times 10^4,1\times 10^5$. For each task, we run $K = 1500$ iterations of estimation and for each time of estimation of gradient, we sample $M=25$ times for all asset sizes except for $1\times 10^5$ where we sample $M=50$ times for better estimation in extra high-dimensional cases. The $\mu_i$ are randomly generated from $\text{Uniform}(-1,1)$ and $\sigma_i$ are randomly generated from $\text{Uniform}(0,0.025)$. 

For Task 2, we consider the news-vendor problem of inventory sizes of $1\times 10^2,1\times 10^3,1\times 10^4,1\times 10^5, \text{and } 1\times 10^6$. For each task, we run $K = 1500$ iterations of estimation and for each time of estimation of gradient, we sample $M=25$ times for all asset sizes except for $1\times 10^6$ where we sample $M=50$ times. The $\mu_i$ are randomly generated from $\text{Uniform}(20,50)$ and $\sigma_i$ are randomly generated from $\text{Uniform}(10,20)$.

For Task 3, we use the synthetic data method from \cite{mukherjee2013parallel} and \cite{byrd2016stochastic}. The synthetic dataset contain $N$ sample each has $n$ binary features. Here the feature size $n$ is the size of the problem. We generate datasets with $50, 500, 1000, 5000$ features, each with sample size $N= 30n$.
The labels are generated by a random linear combination of
the features, and contain 10\% white label noise for binary classification. The stepsize is set as $\alpha_k=\beta/k$. Other parameters are given as $M =25$, $L =10$, $b=50$, $\beta=2$ and $b_H =300$ or $600$. For each round of experiment we run $K = 2000$ iterations.

All our experiments were conducted in a Python 3 environment. For CPU computations, we utilized an AMD Ryzen Threadripper 3970X with 256GB of memory, and for GPU computations, we employed an NVIDIA GeForce RTX 3090 with 24GB of memory, with key parameters are detailed in Table \ref{table:cpu_gpu_comparison}. The CPU and GPU used in the experiment are at comparable market price. 
\begin{table}[htbp]
\centering
\begin{tabular}{c|c|c}
\hline
                            & \textbf{CPU}     & \textbf{GPU}                     \\ \hline
\textbf{Processor}          & AMD Ryzen Threadripper 3970X & NVIDIA GeForce RTX 3090      \\ 
\textbf{Theoretical peak (FP32)} & 108 GFLOPS                        & 35.58 TFLOPS                        \\ 
\textbf{Maximum memory bandwidth} & 172.73 GB/sec                  & 936.2 GB/sec                      \\ \hline
\end{tabular}
\caption{Comparison of CPU and GPU specifications.}
\label{table:cpu_gpu_comparison}
\end{table}

\subsection{Experiment results}
\begin{figure}[htbp]
    \centering
    \includegraphics[scale=0.5]{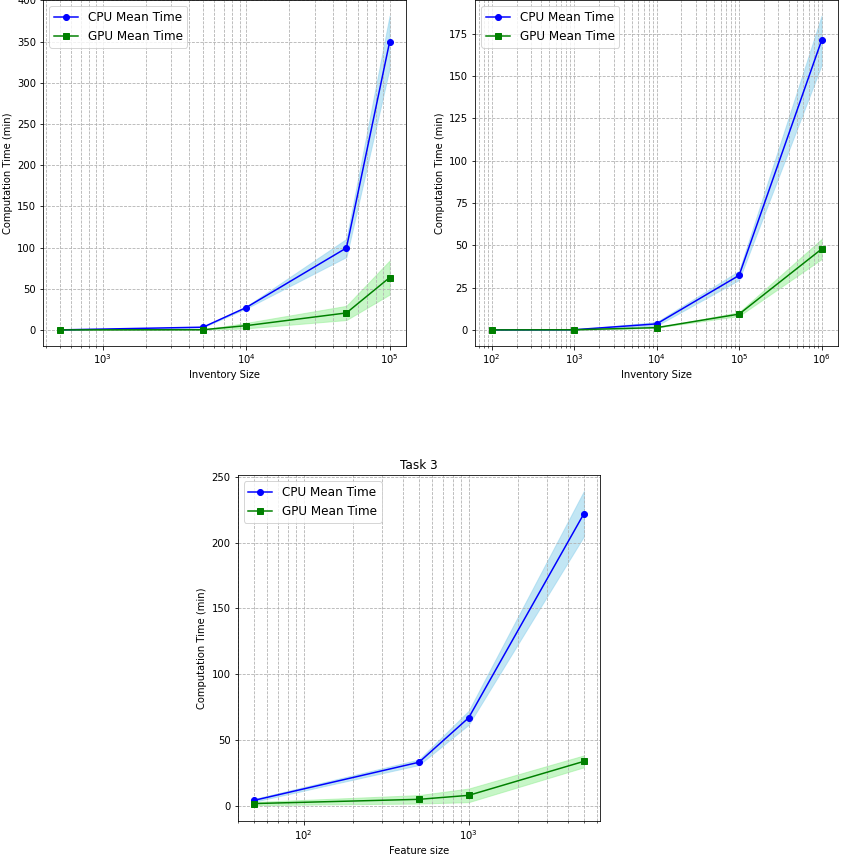}
    \caption{Computation Time for Three tasks with Different Size}
    \label{fig: result}
    \begin{flushleft}
\small
\textbf{Notes:}
This figure demonstrates the computation time and corresponding $\pm 2 \sigma$ confidence interval for three considered tasks of different sizes.
\end{flushleft}
\end{figure}

In Figure \ref{fig: result}, we present the average computation time and the corresponding confidence intervals, defined as plus or minus two standard deviations, for three tasks of varying scales. Additionally, we examine the convergence properties of each task using a selected example size, demonstrating the relative squared error (RSE) of the objective values in comparison to the final objective values.

Figure \ref{fig: result} illustrates that the GPU implementation consistently outperforms in computational time across all three SO tasks. As the problem size increases, the benefits of leveraging GPU implementation for parallel computing and vectorization become increasingly pronounced. For instance, in portfolio optimization tasks involving $10^5$ assets, completing all iterations typically requires around 6 hours. Using GPU technology, however, can reduce this iteration time to approximately 1 hour, thus achieving an acceleration factor of about six. Additionally, Table \ref{table: performance} demonstrates that the same algorithm running on both GPU and CPU achieves nearly identical levels of accuracy at various iteration steps. This similarity in performance is anticipated since, apart from the computation hardware, all other parameters remain the same throughout the process.

\begin{table}[H]
\centering
\begin{tabular}{ccccccc}
\hline
                    & \multicolumn{2}{c}{Asset (5k)} & \multicolumn{2}{c}{Inventory (10k)} & \multicolumn{2}{c}{Classification (1k)} \\
                    & GPU           & CPU          & GPU           & CPU          & GPU           & CPU          \\ \hline
RSE at iteration 50          &   85.07\%           &  83.19\%         & 89.92\%           &  88.73\%         & 72.16\%           &  76.25\%        \\
                    & ($\pm$ 9.74\%)       & ($\pm$ 10.65\%)    & ($\pm$ 7.02\%)       & ($\pm$ 7.33\%)    & ($\pm$ 8.44\%)       & ($\pm$ 7.74\%)    \\
RSE at iteration 100  & 62.41\%      & 63.71\%    & 76.25\%      & 72.93\%    & 51.06\%      & 53.46\%    \\
                 & ($\pm$ 5.46\%) & ($\pm$ 4.86\%) & ($\pm$ 8.49\%) & ($\pm$ 9.45\%) & ($\pm$ 5.92\%) & ($\pm$ 5.10\%) \\
RSE at iteration 500  & 24.07\%          & 25.62\%          & 40.94\%          & 38.52\%          & 31.29\%          & 29.67\%          \\
                 & ($\pm$ 4.97\%)    & ($\pm$ 5.87\%)    & ($\pm$ 8.11\%)    & ($\pm$ 8.53\%)    & ($\pm$ 4.07\%)    & ($\pm$ 5.21\%)    \\
RSE at iteration 1000    & 13.39\%             & 12.93\%          & 20.58\%             & 23.67\%          & 15.59\%             & 16.77\%          \\
                    & ($\pm$ 2.86\%)       & ($\pm$ 3.96\%)    & ($\pm$ 5.78\%)       & ($\pm$ 6.48\%)    & ($\pm$ 4.00\%)       & ($\pm$ 3.71\%)    \\ \hline
\end{tabular}
\caption{Evaluation of Performance on Different Tasks with Adjusted Error Estimates.}
\begin{flushleft}
\small
\textbf{Notes:}
We define the Relative Squared Error (RSE) as $\text{RSE} = \left(\frac{y^{(t)} - y^*}{y^{(t)}}\right)^2 \times 100\%$, where $y^*$ represents the final objective value upon completion of iterations, and $y^{(t)}$ denotes the objective value at the $t^{\text{th}}$ iteration. This table presents the RSE for various optimization tasks: a mean-variance optimization involving 5000 assets, a newsvendor problem with 10,000 products, and binary classification tasks with 1000 features, each assessed under varying iteration step counts within a total of 10,000 iteration steps. We repeat each experiment for 7 repetitions.
\end{flushleft}
\label{table: performance}
\end{table}

\section{Conclusion}
In this paper, we present a preliminary study on employing GPUs to expedite computation across three simulation optimization tasks. We observe that by leveraging the GPU's capabilities for fast vectorization and parallel computing, both first-order and second-order algorithms experience a performance improvement of approximately 3 to 6 times. The relative benefit increases as the problem scale increases. Our study has limitations, including reliance on third-party GPU acceleration packages, which may not fully utilize the computational power of GPUs. Additionally, we have not thoroughly investigated the specific contributions of GPUs at various computational stages. Moreover, our focus is restricted to gradient-based methods, and has not extended to other simulation optimization algorithms.

\bibliographystyle{informs2014}
\bibliography{references}

\end{document}